\documentstyle[12pt]{article}
\newcommand{\ehat}{ \hat U_{\epsilon} }

\newcommand{\define}{ \stackrel{\triangle}{=} }

\def\be{\begin{equation}}
\def\ee{\end{equation}}
\def\ba{\begin{eqnarray}}
\def\ea{\end{eqnarray}}

\begin{document}
\title{\bf Gauge Model With Massive Gravitons  }
\author{{Ning Wu}
\thanks{email address: wuning@mail.ihep.ac.cn}
\\
\\
{\small Institute of High Energy Physics, P.O.Box 918-1,
Beijing 100039, P.R.China}
\thanks{mailing address}}
\maketitle
\vskip 0.8in

~~

\vskip 0.8in

\begin{abstract}
Gauge theory of gravity is formulated based on principle 
of local gauge invariance. Because the model has strict 
local gravitational gauge symmetry, gauge theory of gravity
is a perturbatively renormalizable quantum model. However,
in the original model, all gauge gravitons are massless. 
We want to ask that whether there exists massive gravitons
in Nature? In this paper, we will propose a gauge model 
with massive gravitons. The mass term of gravitational 
gauge field is introduced into the theory without violating 
the strict local gravitational gauge symmetry. Massive
gravitons can be considered to be possible origin of dark 
energy and dark matter in the Universe. 
\\

\end{abstract}

\newpage

\Roman{section}


It is generally believed that four kinds of fundamental
interactions in Nature are gauge interactions and can be
described by gauge field theory. From theoretically point
of view, gauge principle plays a fundamental role
in particle's interactions theory. We know that in 
Yang-Mills theory, if the system has strict local gauge
symmetry, all gauge field must be massless\cite{1}.
But, according to experimental results, some gauge 
bosons are massive\cite{pdg}. Up to now, there are
mainly  two mechanisms 
to introduce the mass term of gauge field. One is
spontaneously symmetry
breaking and Higgs mechanism \cite{5,6,7,8,9,10,11,12},
which is well-known in constructing the standard 
model\cite{2,3,4}. So, in this mechanism, the introduction
of the mass of gauge field violate the local gauge 
symmetry of the system. Another mechanism is proposed by N.Wu 
in literature\cite{13}. In this mechanism, two sets of gauge
field are introduced into the model simultaneously. The 
biggest advantage of this mass generation mechanism 
is that the mass term of gauge fields does not violate 
the local gauge symmetry of the Lagrangian. This model
can also be applied to unified electroweak interactions
and the unified electroweak theory based on this
mechanism has the similar dynamics for low energy
electroweak interactions, but contains no Higgs 
particle\cite{14}, which will become a prospective
theory for unified electroweak interactions if 
Higgs particles do not exist in Nature.   
\\

Gauge theory of gravity is proposed based on gauge 
principle\cite{15,16,17,18}. It is known that gauge 
treatment of gravity was suggested immediately after the
gauge theory birth itself\cite{b1,b2,b3,b4,b5,b6}.
In the traditional gauge treatment of gravity, Lorentz 
group is localized, and the gravitational field is not 
represented by gauge potential, but represented by 
metric field. The theory is non-renormalizable. 
Gauge theory of gravity is proposed in physics picture
of gravity.  The model has strict gravitational
gauge symmetry.  After localization of gravitational 
gauge group, the gravitational field appears as the 
corresponding gauge potential. One advantage of 
gauge theory of gravity is that, in this new theory, 
four different kinds of fundamental interactions in 
Nature can be formulated in the same manner, so
they can be unified together\cite{19,20,21}. 
In this model, we will apply the mass generation mechanism
of gauge field which is preposed in literature \cite{13}
to gravitational gauge theory and to introduce the mass
term of gravitational gauge field withoug violating
the local gauge symmetry of the lagrangian. \\

In gauge theory of gravity, the most fundamental 
quantity is gravitational gauge field. 
In order to introduce mass term of gravitational gauge 
fields, we need two sets of gravitational gauge fields 
simultaneously. Suppose that the first set of gauge fields 
is denoted as $C_{\mu}^{\alpha}$, and the second set of 
gauge fields is denoted as $C_{2 \mu}^{\alpha}$.
For gauge theory of gravity, the symmetry of gravitational
interactions is selected to be gravitational gauge symmetry
and the corresponding gauge group is selected to be
gravitational gauge group. Details on gravitational gauge 
symmetry and gravitational gauge group can be found in
leterature\cite{17,18}. Under gravitational gauge transformation, 
two gravitational gauge fields transform as
\be
C_{\mu}(x) \to  C'_{\mu}(x) =
\ehat (x) C_{\mu} (x) \ehat^{-1} (x)
+ \frac{i}{g} \ehat (x) (\partial_{\mu} \ehat^{-1} (x)),
\label{01}
\ee
\be
C_{2 \mu}(x) \to  C'_{2 \mu}(x) =
\ehat (x) C_{2 \mu} (x) \ehat^{-1} (x)
- \frac{i}{\alpha g} \ehat (x) (\partial_{\mu} \ehat^{-1} (x)),
\label{02}
\ee
where ${\bf \alpha}$ is a constant parameter and 
$\ehat (x)$ is the gravitational gauge transformation
operator\cite{15,16,17,18}
\be
\ehat (x) = E^{-i \epsilon^{\beta} \cdot \hat{P}_{\beta}}.
\label{02001}
\ee
Gravitational gauge fields $C_{\mu}(x)$ and $C_{2 \mu}(x)$
are vectors in gravitational Lie algebra\cite{17,18}, so they
can be expanded as
\be
C_{\mu}(x) = C_{\mu}^{\alpha}(x) \hat{P}_{\alpha}, 
\label{0201}
\ee
\be
C_{2 \mu}(x) = C_{ 2 \mu}^{\alpha}(x) \hat{P}_{\alpha}.
\label{0202}
\ee
Correspondingly, there are two gauge covariant derivatives,
\be
D_{\mu} = \partial_{\mu} - i g C_{\mu} (x),
\label{03}
\ee
\be
D_{2 \mu} = \partial_{\mu} + i\alpha g C_{2 \mu} (x),
\label{04}
\ee
and two different strengths of gauge fields,
\be
F_{\mu \nu} = \frac{1}{-i g}
\lbrack D_{\mu} ~~,~~ D_{\nu} \rbrack,
\label{05}
\ee
\be
F_{2 \mu \nu} = \frac{1}{i \alpha g}
\lbrack D_{2 \mu} ~~,~~ D_{2 \nu} \rbrack.
\label{06}
\ee
The explicit forms of field strengths are
\be
F_{\mu \nu} = \partial_{\mu} C_{\nu}(x)
- \partial_{\nu} C_{\mu}(x)
- i g C_{\mu}(x) C_{\nu}(x)
+ i g  C_{\nu}(x) C_{\mu}(x),
\label{07}
\ee
\be
F_{2 \mu \nu} = \partial_{\mu} C_{2 \nu}(x)
- \partial_{\nu} C_{2 \mu}(x)
+ i \alpha g C_{2 \mu}(x) C_{2 \nu}(x)
- i \alpha g  C_{2 \nu}(x) C_{2 \mu}(x).
\label{08}
\ee
Field strengths $F_{\mu \nu}$ and $F_{2 \mu \nu}$
are also vectors in gravitational Lie algrbra, so
they can be expanded in terms of generators of 
gravitational gauge group
\be
F_{\mu \nu} = F_{\mu \nu}^{\gamma} \hat{P}_{\gamma},
\label{09}
\ee
\be
F_{2 \mu \nu} = F_{2 \mu \nu}^{\gamma} \hat{P}_{\gamma},
\label{10}
\ee
where $F_{\mu \nu}^{\gamma}$ and $F_{2\mu \nu}^{\gamma}$
are component field strengths. 
The explicit forms of component strengths are
\be
F_{\mu \nu}^{\gamma} = \partial_{\mu} C_{\nu}^{\gamma}
- \partial_{\nu} C_{\mu}^{\gamma}
- g C_{\mu}^{\beta} \partial_{\beta} C_{\nu}^{\gamma}
+ g C_{\nu}^{\beta} \partial_{\beta} C_{\mu}^{\gamma},
\label{11}
\ee
\be
F_{2 \mu \nu}^{\gamma} = \partial_{\mu} C_{2 \nu}^{\gamma}
- \partial_{\nu} C_{2 \mu}^{\gamma}
+ \alpha g C_{2 \mu}^{\beta} \partial_{\beta} C_{2 \nu}^{\gamma}
- \alpha g C_{2 \nu}^{\beta} \partial_{\beta} C_{2 \mu}^{\gamma}.
\label{12}
\ee
Using eq.(\ref{01}-\ref{02}), 
we can obtain the following transformation properties,
\be
D_{\mu} (x) \to D'_{\mu} (x)
= \ehat D_{\mu} (x) \ehat^{-1},
\label{13}
\ee
\be
D_{2 \mu} (x) \to D'_{2 \mu} (x)
= \ehat D_{2 \mu} (x) \ehat^{-1},
\label{14}
\ee
\be
F_{\mu \nu} \to F'_{\mu \nu} =
\ehat F_{\mu \nu} \ehat^{-1},
\label{15}
\ee
\be
F_{2 \mu \nu} \to F'_{2 \mu \nu} =
\ehat F_{2 \mu \nu} \ehat^{-1},
\label{16}
\ee
\be
(C_{\mu} + \alpha C_{2 \mu})
\to (C'_{\mu} + \alpha C'_{2 \mu}) =
\ehat  (C_{\mu} + \alpha C_{2 \mu}) \ehat^{-1}.
\label{17}
\ee
\\

Matrix $G$ is an important quantity in gauge theory
of gravity, whose definition is\cite{15,16,17,18}
\be \label{19}
G = (G_{\mu}^{\alpha}) = ( \delta_{\mu}^{\alpha} - g C_{\mu}^{\alpha} ).
\ee
Its inverse matrix is denoted as $G^{-1}$,
\be \label{20}
G^{-1} = \frac{1}{I - gC} = (G^{-1 \mu}_{\alpha}).
\ee
They satisfy the following relations,
\be \label{21}
G_{\mu}^{\alpha} G^{-1 \nu}_{\alpha} = \delta_{\mu}^{\nu},
\ee
\be \label{22}
G_{\beta}^{-1 \mu} G^{ \alpha}_{\mu} = \delta_{\beta}^{\alpha}.
\ee
It can be proved that
\be \label{23}
D_{\mu}= G_{\mu}^{\alpha} \partial_{\alpha}.
\ee
\\

In order to construct a gravitational gauge invariant
lagrangian, $J(C)$ is an important factor. In this paper,
it will select to be\cite{15,16,17,18}
\be \label{24}
J(C) = \sqrt{- {\rm det}( g_{\alpha \beta} ) },
\ee
where
\be \label{25}
g_{\alpha \beta} \define \eta_{\mu \nu}
(G^{-1})_{\alpha}^{\mu} (G^{-1})_{\beta}^{\nu}.
\ee
\\

The Lagrangian of the system is
\be \label{26}
\begin{array}{rcl}
{\cal L}_0 &=& - \frac{1}{4} \eta^{\mu \rho} \eta^{\nu \sigma}
g_{\beta\gamma}
F_{\mu \nu}^{\beta} F_{\rho \sigma}^{\gamma}
- \frac{1}{4} \eta^{\mu \rho} \eta^{\nu \sigma}
g_{\beta\gamma}
F_{2 \mu \nu}^{\beta} F_{2 \rho \sigma}^{\gamma} \\
&&\\
&& - \frac{m^2}{2(1+\alpha^2)}
\eta^{\mu \nu}  g_{\beta \gamma}
(C_{\mu}^{\beta}  + \alpha C_{2 \mu}^{\beta})
(C_{\nu}^{\gamma} + \alpha C_{2 \nu}^{\gamma}),
\end{array}
\ee
where $m$ is a constant mass parameter. 
The  action  is defined by 
\be \label{27}
S = \int {\rm d}^4 x J(C) {\cal L}_0.
\ee
Using relations (\ref{13} - \ref{17}) and the following
relation
\be \label{28}
\int {\rm d}^4 x J \cdot (\ehat f(x))
= \int {\rm d}^4 x f(x)
\ee
where 
$J ={\rm det} ( \frac{ \partial(x - \epsilon(x))^{\beta}} 
{\partial x^{\alpha}}) $ 
is the Jacobian of the corresponding transformations 
and $f(x)$ is an arbitrary function\cite{15,16,17,18}, 
it is easy to prove that the action $S$ has local 
gravitational gauge symmetry. \\

From eq.(\ref{26}), we can  see that there is mass 
term of gravitational gauge fields. In order to obtain 
the eigenstates of mass matrix, a rotation is needed
\be
C_{3 \mu} = {\rm cos}\theta C_{\mu}
+ {\rm sin}\theta C_{2 \mu},
\label{29}
\ee
\be
C_{4 \mu} = - {\rm sin}\theta C_{\mu}
+ {\rm cos}\theta C_{2 \mu},
\label{30}
\ee
where $\theta$ is given by
\be
{\rm cos}\theta = 1 / \sqrt{1 + \alpha^2 },~~~~
{\rm sin}\theta = \alpha / \sqrt{1 + \alpha^2 }.
\label{31}
\ee
After these transformations, the lagrangian ${\cal L}_0$
will becomes
\be
{\cal L}_0 = - \frac{1}{4} \eta^{\mu \rho} \eta^{\nu \sigma}
g_{\beta\gamma}
F_{30 \mu \nu}^{\beta} F_{30 \rho \sigma}^{\gamma}
- \frac{1}{4} \eta^{\mu \rho} \eta^{\nu \sigma}
g_{\beta\gamma}
F_{40 \mu \nu}^{\beta} F_{40 \rho \sigma}^{\gamma} 
- \frac{m^2}{2} \eta^{\mu \nu} g_{\beta\gamma}
 C_{3 \mu}^{\beta}   C_{3 \nu}^{\gamma} 
+{\cal L}_I ,
\label{32}
\ee
where ${\cal L}_I$ contains all interaction terms of 
gravitational gauge fields and 
\be  \label{33}
F_{30 \mu \nu}^{\alpha} =
\partial_{\mu} C_{3 \nu}^{\alpha}
- \partial_{\nu} C_{3 \mu}^{\alpha},
\ee
\be  \label{34}
F_{40 \mu \nu}^{\alpha} =
\partial_{\mu} C_{4 \nu}^{\alpha}
- \partial_{\nu} C_{4 \mu}^{\alpha}.
\ee
So, the gravitational gauge field $C_{3 \mu}$ is massive whose
mass is $m$ while gravitational gauge field $C_{4 \mu}$ is kept
massless. \\

Gravitational gauge field $C_{3 \mu}$ is massive. If its
mass is very large, it will 
have no contribution to the long range gravitational force.
Long range gravitational force completely comes from the
contribution of gravitational gauge field $C_{4 \mu}$
and obeys inverse square law.
But, if the mass term of gravitational gauge field 
$C_{3 \mu}$ is extremely small, gravitational gauge field
$C_{3 \mu}$ will also contribute some to the
middle range gravitational force. Its force range $l$ 
is about
\be  \label{35}
l \sim \frac{hc}{m},
\ee
where $h$ is the plank constant and $c$ is the
speed of light. So, if $m$ is about
$2 \times 10^{-7}$ eV, its force range $l$ will be
about one meter. The gravitational force carried by
gravitatioanl gauge field  $C_{3 \mu}$ decreases
exponentially. In this case, the inverse square law
of gravitational force in middle range will be violated. 
But long range gravitational force still obey 
inverse square law. 
\\

The existence of massive gravitons in Nature is very 
important for cosmology. Because the coupling constant 
for gravitational interactions is weak and  the massive 
gravitons only take part in gravitational interactions,
the massive gravitons must be a relative stable particle 
in Nature and they have very weak coupling with 
ordinary matter. They have no direct
electromagnetic interactions with photons, so they radiate
no photons. Because they take part in gravitational 
interactions, they will contribute some to the average
energy density of the Universe, and affect the global structure
and evolution of the Universe. In a ward, they can not
be seen, but they are heavy and affect motions of ordianry
celestial objects and global structure and evolution 
of our Universe. So, if they indeed exist, it 
must be dark matter or dark energy in the Universe. 
According to recent results of cosmological observations,
especially from Cosmic Microwave Bakground (CMB)
temperature anisotropies, our Universe is spacially flat,
consists of mainly dark matter and dark energy\cite{22}. 
A natural origin for dark energy and dark 
energy is to regard them as gravitons.  
\\

If massive gravitons exist in Nature, it will be 
extremely hard for us to directly detect them
in experiments, for they can not be seen in any experiments
and they have very weak coupling with ordianry matter. 
For a single massive graviton, because its mass is very
small, its gravitational interactions with any experimental
instruments is almost zero, it will leave no signal 
to any experimental instruments. So, the only possibility
to prove the existence of mass graviton is to find evidence
in astrophysical and cosmological phenomenon. 
\\

Because Higgs mechanism violates local gauge symmetry of
the system and we know that gravitational gauge symmetry
is a strict symmetry, Higgs mechanism can not be used to
introduce the mass term of gravitational gauge fields. \\

\end{document}